\documentclass[aps, prd, groupedaddress, preprint, tightenlines,  eqsecnum, nofootinbib, showpacs]{revtex4}
\usepackage{epsfig}
\usepackage{setspace}
\usepackage{graphicx}
\usepackage{leqno}
\usepackage{cancel}
\usepackage{amsmath}
\usepackage{amsfonts}
\usepackage{amssymb}
\usepackage{enumerate}

\usepackage{listings}
\usepackage{axodraw, float, slashed, graphicx, amssymb, amsmath}
\usepackage[usenames, dvipsnames]{xcolor}
\usepackage{booktabs}

\usepackage[margin=2.2cm, a4paper, includefoot]{geometry}

\def\Li{\mathop{\hbox{\rm Li}}\nolimits}
\def\e{\epsilon}

\def\spa#1.#2{\left\langle#1\,#2\right\rangle}
\def\spb#1.#2{\left[#1\,#2\right]}
\def\la{\langle}
\def\ra{\rangle}

\def\NeqFour{\Neq4}

\newcommand{\tree}{\text{tree}}
\newcommand{\Neq}[1]{\mathcal{N} = #1}

\DeclareMathOperator{\tr}{\mathrm{tr}}

\def\st{I^{(1)}}

\newcommand\figref[1]{fig.~\ref{#1}}
\def\eps{\epsilon}

\def\be{\begin{equation}}
\def\ee{\end{equation}}

\def\Log{{\rm Log}}
\begin{document}

\hfill\today

\title{The two-loop $n$-point all-plus helicity amplitude} 

\author{David~C.~Dunbar, Guy R. Jehu and Warren~B.~Perkins}

\affiliation{
College of Science, \\
Swansea University, \\
Swansea, SA2 8PP, UK\\
\today
}

\begin{abstract}
We propose a compact analytic expression for the polylogarithmic part of the $n$-point two-loop all-plus helicity amplitude in gauge theory.   
\end{abstract}


\maketitle

\section{Introduction}

Computing perturbative scattering amplitudes is a key challenge in quantum field theory both for comparing theories with experiment and for
understanding the symmetries and consistency of theories. 
Explicit analytic expressions for scattering amplitudes have proved particularly useful in understanding the behaviour and symmetries of the underlying theory.  
Calculating amplitudes from their singular structures is  an extremely powerful tool which has a long history~\cite{Eden} and has seen tremendous development based 
on techniques including unitarity~\cite{Bern:1994zx,Bern:1994cg} and on-shell recursion~\cite{Britto:2005fq}.

Recently the leading in color component of the two-loop all-plus five-point amplitude has been computed in QCD~\cite{Badger:2013gxa, Badger:2015lda} using $d$-dimensional unitarity techniques.  Subsequently this amplitude was presented in a very elegant and compact form~\cite{Gehrmann:2015bfy}.  In this form the amplitude consists of a piece driven by the infra-red (IR) and ultra-violet (UV) singular structure of the amplitude and a ``remainder'' piece.

In ref.~\cite{Dunbar:2016aux} it was 
demonstrated how this form can be generated using a combination of four-dimensional unitarity and (augmented) 
recursion which provides an understanding of the simplicity of the amplitude. 
In this article  we propose an expression for the polylogarithms in the leading-in-color part of the $n$-point all-plus amplitude. This proposal is
based on collinear limits and unitarity.

Using the conventions of ref.~\cite{Gehrmann:2015bfy} the leading in color component of the amplitude can be expressed, 
\begin{align}
{\mathcal A}_{n}(1^+, 2^+ ,\cdots ,n^+) |_{\rm leading \; color}=& g^{n-2}  \sum_{L \ge 1} \left( g^2 N c_{\Gamma} \right)^L  \nonumber\\
&\hspace{-4cm} \times \sum_{\sigma \in S_{n}/Z_{n}} {\rm tr}(T^{a_{\sigma(1)}} T^{a_{\sigma(2)}} \cdots T^{a_{\sigma(n)}}) 
A^{(L)}_{n}(\sigma(1)^{+} , \sigma(2)^{+} , \cdots ,\sigma(n)^{+})
\end{align}
and the result we present is for the color-stripped two-loop amplitude 
$A^{(2)}_{n}(1^+,2^+,\cdots ,n^+)$.\footnote{The factor $c_{\Gamma}$
is defined as $\Gamma(1+\epsilon)\Gamma^2(1-\epsilon)/\Gamma(1-2\epsilon)/(4\pi)^{2-\epsilon}$.  
$S_n/Z_n$ are the cyclically-distinguishable permutations of the $n$-legs and $T^{a_i}$ are the color-matrices of $SU(N)$.}

The IR and UV behaviour of this amplitude is very-well specified~\cite{Catani:1998bh} and it can be  split into singular terms plus a finite remainder function,
$F_n^{(2)}$, which is to be determined:
\begin{align}\label{definitionremainder}
A^{(2)}_{n} =& A^{(1)}_{n}\st_n +  \;F^{(2)}_{n}  + {\mathcal O}(\eps)\,,
\end{align}
where 
\begin{align}\label{definitionremainderI}
\st_{n} =& \left[ - \sum_{i=1}^{n} \frac{1}{\epsilon^2} \left(\frac{\mu^2}{-s_{i,i+1}}\right)^{\epsilon} 
+\frac{n\pi^2}{12} \right] \; . 
\end{align}
Since this one-loop amplitude is finite there are no $\epsilon^{-1}$ terms~\cite{Catani:1998bh}. 
In this equation $A^{(1)}_{n}$ is the all-$\epsilon$ form 
of the one-loop amplitude~\cite{Bern:1993qk,Bern:1996ja}. 
Although the one-loop amplitude is rational  to ${\cal O}(\epsilon^0)$,  the all-$\epsilon$ expression contains polylogarithms which, when combined with the 
$\eps^{-2}$ factor generate finite polylogarithms in the two-loop amplitude.  $F_n^{(2)}$ contains further polylogarithmic terms, $ P_n^{(2)}$ 
and rational terms $R_n^{(2)}$:
\begin{equation}
F_n^{(2)} = P_n^{(2)}+R_n^{(2)} \; . 
\end{equation}

The ansatz for $P_n^{(2)}$ is in terms of the functions
\begin{align}
F^{2m}[ S,T, K_2^2, K_4^2] = 
&\Li_2[1-\frac{K_2^2}{S}]+
\Li_2[1-\frac{K_2^2}{T}]+
\Li_2[1-\frac{K_4^2}{S}]
\notag \\
+&
\Li_2[1-\frac{K_4^2}{T}]-
\Li_2[1-\frac{K_2^2K_4^2}{ST}]+
\Log^2(S/T)/2 \; . 
\end{align}
As the notation suggests, these are related to the one-loop box integral functions with two massive (and two massless) legs.
The function corresponds to a box integral function where the momenta $K_2$ and $K_4$ are non-null and $S=(k_1+K_2)^2$ and $T=(K_2+k_3)^2$.   
This combination of dilogarithms can be regarded as either a) the $D=4$ integrals truncated to remove the IR divergent 
terms or b) the $D=8$ integrals~\cite{BrittoUnitarity,Bidder:2005ri}.  
The function is smooth in the limit $K_2^2\longrightarrow 0$,
\begin{equation}
F^{2m}[ S,T, 0, K_4^2] = 
\Li_2[1-\frac{K_4^2}{S}]+
\Li_2[1-\frac{K_4^2}{T}]
+\Log^2(S/T)/2
+\frac{\pi^2}{6}
\end{equation}
which is the one-mass function, $F^{1m}[S,T,K_4^2]$.

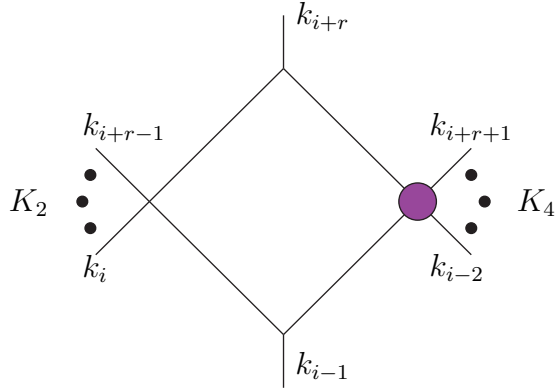
\begin{figure}[h]
\centerline{
    \begin{picture}(170,150)(0,-20)    
   \Line(75,0)(25,50)
   \Line(75,0)(125,50)
   \Line(125,50)(75,100)
   \Line(25,50)(75,100)
   \Line(75,0)(75,-20)
   \Line(75,100)(75,120)
   \Line(25,50)(5,30)
    \Line(25,50)(5,70)
    \Line(125,50)(145,70)
    \Line(125,50)(145,30)
     \Text(90,-12)[c]{$k_{i-1}$}  
    \Text(90, 120)[c]{$k_{i+r}$}  
     \Text(-20,50)[c]{$K_2$}  
          \Text(0,25)[l]{$k_i$}  
           \Text(0,78)[l]{$k_{i+r-1}$}  
           \Text(130,78)[l]{$k_{i+r+1}$}  
           \Text(130,25)[l]{$k_{i-2}$}  
     \Text( 170, 50)[c]{$K_4$}  
   \CCirc(125,50){7}{Black}{Purple} 
      \CCirc(0,50){2}{Black}{Black} 
      \CCirc(3,60){2}{Black}{Black} 
       \CCirc(3,40){2}{Black}{Black} 
           \CCirc(150,50){2}{Black}{Black} 
                          \CCirc(145,60){2}{Black}{Black} 
                               \CCirc(145,40){2}{Black}{Black} 
    \end{picture} 
    }
    \caption{ Diagrammatic representation of the functions  $F^{2m}_{n:r,i}$. 
   The functions are symmetric between $K_2$ and $K_4$ but their coefficients are not.  
   The summation is over all such functions including the case when 
   $K_2$ is a single leg ($r=1$) but leg $K_4$ must contain at least two legs 
   (indicated by a solid disc).}
    \label{fig:boxdefs}
\end{figure}

$P_n^{(2)}$ is obtained by summing over all possible $F^{2m}$ including the degenerate cases when $K_2$  corresponds to a single leg and $F^{2m}$ reduces to $F^{1m}$. Leg $K_4$ must have at least two external legs.  Defining the kinematic invariants $t_i^{[r]}$
\begin{equation}
  t_i^{[r]} =(k_i+k_{i+1}+\cdots +k_{i+r-1})^2
\end{equation}   
then
\begin{equation}
F^{2m}_{n:r,i}=F^{2m}[ t_{i-1}^{[r+1]},t_i^{[r+1]},  t_{i}^{[r]}, t_{i+r+1}^{[n-r-2}] 
\end{equation}
which is shown diagramatically in \figref{fig:boxdefs}.
The case $r=1$ is the one-mass case $F^{1m}_{n:i}\equiv F^{2m}_{n:1,i}$.

The ansatz for $P_n^{(2)}$ is 
\begin{equation}
P_n^{(2)}   =  -{ i  \over 
3 \spa1.2\spa2.3\spa3.4\cdots \spa{n}.1   }\sum_{i=1}^n  \sum_{r=1}^{n-4} c_{r,i}  F^{2m}_{n:r,i}
\end{equation}
where the coefficient is 
\begin{align}
c_{r,i}= \left( 
\sum_{a<b<c<d \in K_4}  \tr_{-} [abcd]-\sum_{a<b<c \in K_4} \tr_{-}[abc  K_4]  +\sum_{a<b \in K_4 } { \la {i-1} |K_4  a b K_4 | {i+r} \ra  \over \spa{{i-1}}.{{i+r}} }
\right)\; 
\end{align}
where 
\begin{align}
\tr_{-} [abcd] \equiv \spa{a}.b\spb{b}.c\spa{c}.d\spb{d}.a \; ,
\end{align}
$K_4$ is the set  $\{ i+r+1, \cdots , i-2\}$ with a cyclic definition of indices and inequality refers to ordering within the set.

Note that $P_n^{(2)}$ has transcendentality two (w.r.t polylogarithms) as do one-loop amplitudes although the full amplitude, $A_n^{(2)}$, 
has higher transcendentality.  For $n=4$, $P_4^{(2)}=0$ and the remainder function is purely rational~\cite{Bern:2002tk}.

\section{Collinear Limits}  

We consider the collinear limit of the amplitude as an important consistency test and to illustrate some key features.  The collinear limit
occurs when adjacent momenta $k_a$ and $k_{a+1}$ become collinear,
\begin{equation}
k_a \longrightarrow z \times K , \;\;\ k_{a+1} \longrightarrow (1-z) \times K = \bar z K \; .
\end{equation}
In this limit, amplitudes factorise as 
\begin{equation}
A^{(L)}_n (\cdots ,k_a^{h} , k_{a+1}^{h'}, \cdots )
\longrightarrow  
\sum_{L_s,h''}    S^{hh', (L_s) }_{-h''} \times A^{(L-L_s)}_{n-1} (\cdots ,K^{h''}, \cdots )\; ,
\end{equation}
where $S^{hh',(L_s)}_{-h''}$ are the various splitting functions.   For the all-plus  amplitude the tree amplitude vanishes for both choices of $h''$ and
\begin{equation}
A^{(2)}_{n}(\cdots k_a^{+} , k_{a+1}^{+} \cdots )
\longrightarrow  
 S^{++, \tree }_{-} \times A^{(2)}_{n-1} (\cdots , K^{+}, \cdots )
 +\sum_{h=\pm} S^{++,(1)}_{-h}\times  A^{(1)}_{n-1} (\cdots ,K^{h}, \cdots ) \; .
\end{equation}
The term with $h=-$ is purely rational and is irrelevent for the polylogarithmic term. 

The first important proposal is that we use an all-$\epsilon$ form of the one-loop amplitude  to ensure  that {\it to all orders in $\epsilon$},
\begin{equation}
A^{(1)}_n \longrightarrow S^{++,\tree}_- \times A^{(1)}_{n-1} \; .
\end{equation}
 With this,  we only need the order $\epsilon^0$ form of the one-loop amplitude to check the collinear limit.

In the collinear limit, 
\begin{equation}
\st_n \longrightarrow \st_{n-1} +r^{++}_- +\Delta \; ,
\end{equation}
where~\cite{Bern:1994zx}  
\begin{equation}
r^{++}_-
=
 - {1\over\epsilon^2}\left( {\mu^2\over z\bar{z}(-s_{a,a+1})}\right)^{\epsilon}
 + 2 \ln z\,\ln \bar{z}+ {1\over3} z\bar{z}- {\pi^2\over6}
\end{equation}
and 
\begin{equation}
\Delta=\log(\frac{-s_{a,a+1}}{\mu^2})\log(z\bar z)-\log(\frac{-s_{a-1,a}}{\mu^2})\log(z)-\log(\frac{-s_{a+1,a+2}}{\mu^2})\log(\bar z)
-\log(z)\log(\bar z) -\frac{1}{3} z \bar z +\frac{\pi^2}{4} \; .
\label{eq:Delta}
\end{equation}
The combination  $S^{++,\tree}_-  r^{++}_{-}$ is the one-loop splitting function.
Consequently, 
\begin{align}
A^{(1)}_n \times \st_n &\longrightarrow 
S^{++,\tree}_- A^{(1)}_{n-1}   \left(   \st_{n-1} + r^{++}_- +\Delta \right)
\notag \\
&=  S^{++,\tree}_- \bigl( A^{(1)}_{n-1} \st_{n-1} \bigr)
+\left(S^{++,\tree}_- r^{++}_- \right) A^{(1)}_{n-1}
+S^{++,\tree}_- A^{(1)}_{n-1} \Delta \; .
\notag \\
&=  S^{++,\tree}_- \bigl( A^{(1)}_{n-1} \st_{n-1} \bigr)
+S^{++,(1)}_- A^{(1)}_{n-1}
+S^{++,\tree}_- A^{(1)}_{n-1} \Delta \; .
\end{align}
In the last term, $S^{++,\tree}_- A^{(1)}_{n-1} \Delta$, we need only keep the one-loop amplitude to order $\epsilon^0$.  
Consequently we require
\begin{equation}
P_n^{(2)} \longrightarrow S^{++,\rm tree}_{-}  P^{(2)}_{n-1} -S^{++,\rm tree}_{-} A^{(1)}_{n-1} \Delta'
\label{eq:collinearrequirement} 
\end{equation}
where $\Delta'$ is the non-rational part of $\Delta$ of eq.~(\ref{eq:Delta}).

The overall pre-factor of 
\begin{align}
{i \over 3 \spa1.2\spa2.3\spa3.4\cdots \spa{n}.1   } &\longrightarrow
{1 \over z \bar z \spa{a}.{a+1} } \times   {i\over 3 \spa1.2 \cdots  \spa{a-1}.{K} \spa{K}.{a+2}   \cdots \spa{n}.1}
\notag \\
& =  S^{++,\rm tree}_{-} \times {i\over 3 \spa1.2 \cdots \spa{a-1}.{K} \spa{K}.{a+2}  \cdots \spa{n}.1}
\end{align}
which is the tree splitting function times the descendant $n-1$-point pre-factor.  Consequently we require 
\begin{equation}
\sum_{i=1}^n  \sum_{r=1}^{n-4} c_{r,i}  F^{2m}_{n:r,i}\longrightarrow \sum_{i=1}^{n-1}  \sum_{r=1}^{n-5} c_{r,i}  F^{2m}_{n-1:r,i}-{\bar A}^{(1)}_{n-1} \Delta'
\end{equation}
where $\bar A^{(1)}_{n-1}$ is the one-loop amplitude divided by the prefactor.

\begin{figure}[H]
    \begin{picture}(450,100)(0,0)  
\CCirc(25,50){2}{Black}{Black}
\CCirc(30,57){2}{Black}{Black}
\CCirc(30,43){2}{Black}{Black}
\CCirc(165,50){2}{Black}{Black}
\CCirc(170,57){2}{Black}{Black}
\CCirc(170,43){2}{Black}{Black}
\CCirc(305,50){2}{Black}{Black}
\CCirc(310,57){2}{Black}{Black}
\CCirc(310,43){2}{Black}{Black}
\Line(50,50)(75,75)
\Line(50,50)(75,25)
\Line(75,25)(100,50)
\Line(75,75)(100,50)
\Line(75,75)(75,100)
\Line(75,25)(75,0)
\Line(50,50)(25,30)
\Line(50,50)(25,70)
\Line(100,50)(125,70)
\Line(100,50)(125,30)
\Line(190,50)(215,75)
\Line(190,50)(215,25)
\Line(215,25)(240,50)
\Line(215,75)(240,50)
\Line(215,75)(215,100)
\Line(215,25)(215,0)
\Line(190,50)(165,30)
\Line(190,50)(165,70)
\Line(240,50)(265,70)
\Line(240,50)(265,30)
\Line(330,50)(355,75)
\Line(330,50)(355,25)
\Line(355,25)(380,50)
\Line(355,75)(380,50)
\Line(355,75)(355,100)
\Line(355,25)(355,0)
\Line(330,50)(305,30)
\Line(330,50)(305,70)
\Line(380,50)(405,70)
\Line(380,50)(405,30)
\Text(130,75)[c]{$k_a$}
\Text(125,25)[c]{$k_{a+1}$}
\Text(230,0)[c]{$k_{a+1}$}
\Text(260,25)[c]{$k_{a}$}
\Text(370,100)[c]{$k_{a}$}
\Text(415,78)[c]{$k_{a+1}$}
\SetColor{Green}
\SetWidth{2}
\Line(100,50)(125,30)
\Line(100,50)(125,70)
\Line(215,25)(215,0)
\Line(240,50)(265,30)
\Line(355,75)(355,100)
\Line(380,50)(405,70)
\SetWidth{1}
\CCirc(100,50){5}{Black}{Purple}
\CCirc(240,50){5}{Black}{Purple}
\CCirc(380,50){5}{Black}{Purple}
\end{picture}    
\caption{Functions whose coefficients vanish in the collinear limit}
\label{fig:vanishing}
\end{figure}
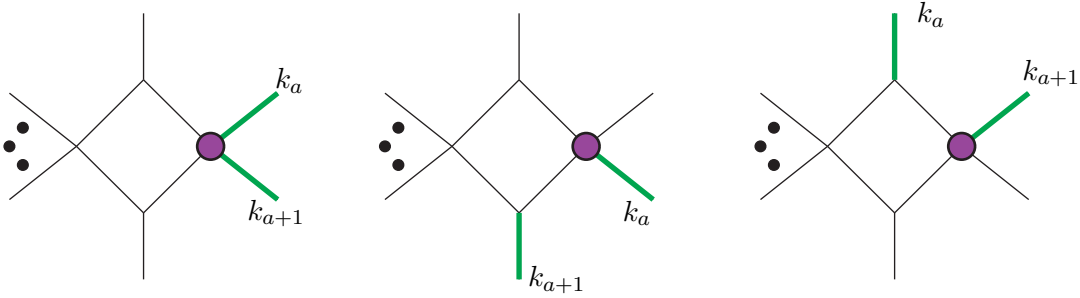

Each term in $P^{(2)}_{n-1}$ where leg $K$ is within either $K_2$ or $K_4$ arises directly from a single term in $P^{(2)}_{n}$ with legs $a$ and $a+1$ both
in  $K_2$ or $K_4$ respectively. In these cases the functions have a smooth limit:
\begin{align}
&F^{2m}_{n:r,i}  \longrightarrow  F^{2m}_{n-1:r-1,i}  & a,a+1 \in K_2 
\notag \\
&F^{2m}_{n:r,i}  \longrightarrow  F^{2m}_{n-1:r,i}     & a,a+1 \in K_4 
\end{align}
while the corresponding coefficients behave as
\begin{align}
& c_{r,i} \longrightarrow c_{r-1,i}  & a,a+1 \in K_2 
\notag \\
&c_{r,i} \longrightarrow c_{r,i} & a,a+1 \in K_4, r\neq n-4 
\end{align}
When $a$ and $a+1$ are the only two legs within $K_4$ (i.e. $r=n-4$),
\begin{equation}
c_{n-4,a+3} =   {\la  a-1|  K_4 (a+1) a K_4 | a+2\ra \over \spa{a-1}.{a+2} }
= { \spa{a-1}.{a} \spb{a}.{a+1}^2  \spa{a}.{a+1} \spa{a+1}.{a+2} \over \spa{a-1}.{a+2}  }
\end{equation}
which vanishes in the collinear limit and there is no contribution from this function . 
This configuration is illustrated in the first part of~\figref{fig:vanishing}.

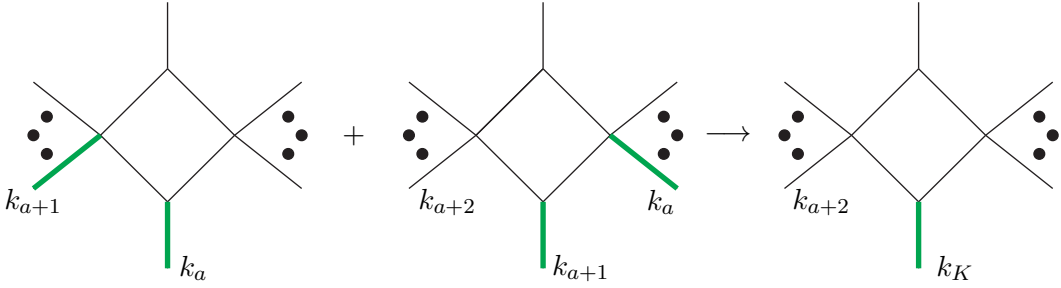
\begin{figure}[H]
    \begin{picture}(450,100)(0,0)  
\CCirc(25,50){2}{Black}{Black} 
\CCirc(30,57){2}{Black}{Black} 
\CCirc(30,43){2}{Black}{Black} 
\CCirc(125,50){2}{Black}{Black} 
\CCirc(120,57){2}{Black}{Black} 
\CCirc(120,43){2}{Black}{Black} 
 \Line(50,50)(75,75)
\Line(50,50)(75,25)
\Line(75,25)(100,50)
\Line(75,75)(100,50)
\Line(75,75)(75,100)
\Line(75,25)(75,0)
\Line(50,50)(25,30)
\Line(50,50)(25,70)
\Line(100,50)(125,70)
\Line(100,50)(125,30)
\Text(145,50)[c]{$+$}
\CCirc(165,50){2}{Black}{Black} 
\CCirc(170,57){2}{Black}{Black} 
\CCirc(170,43){2}{Black}{Black} 
\CCirc(265,50){2}{Black}{Black} 
\CCirc(260,57){2}{Black}{Black} 
\CCirc(260,43){2}{Black}{Black} 
\Line(190,50)(215,75)
\Line(190,50)(215,75)
\Line(190,50)(215,25)
\Line(215,25)(240,50)
\Line(215,75)(240,50)
\Line(215,75)(215,100)
\Line(215,25)(215,0)
\Line(190,50)(165,30)
\Line(190,50)(165,70)
\Line(240,50)(265,70)
\Line(240,50)(265,30)
\Text(285,50)[c]{$\longrightarrow$}
\CCirc(305,50){2}{Black}{Black}
\CCirc(310,57){2}{Black}{Black}
\CCirc(310,43){2}{Black}{Black}
\CCirc(405,50){2}{Black}{Black}
\CCirc(400,57){2}{Black}{Black}
\CCirc(400,43){2}{Black}{Black}
\Line(330,50)(355,75)
\Line(330,50)(355,25)
\Line(355,25)(380,50)
\Line(355,75)(380,50)
\Line(355,75)(355,100)
\Line(355,25)(355,0)
\Line(330,50)(305,30)
\Line(330,50)(305,70)
\Line(380,50)(405,70)
\Line(380,50)(405,30)
\Text(85,0)[c]{$k_a$}
\Text(25,25)[c]{$k_{a+1}$}
\Text(230,0)[c]{$k_{a+1}$}
\Text(180,25)[c]{$k_{a+2}$}
\Text(320,25)[c]{$k_{a+2}$}
\Text(260,25)[c]{$k_{a}$}
\Text(370,0)[c]{$k_{K}$}
\SetColor{Green}
\SetWidth{2}
\Line(50,50)(25,30)
\Line(75,25)(75,0)
\Line(215,25)(215,0)
\Line(240,50)(265,30)
\Line(355,25)(355,0)
\end{picture}    
\caption{Pictorial representation of the identity amongst the $F$-functions..}
\label{fig:identity}
\end{figure}

When one of the massless legs is $a$ or $a+1$  we use the identity
\begin{equation}
F^{2m}_{n:r+1,a+1}+F^{2m}_{n:r,a+2}
\longrightarrow 
F^{2m}_{n-1:r,a+2}
\label{eq:twomassidentity}
\end{equation}
which is shown diagrammatically on \figref{fig:identity}.
This identity follows from Abel's identity.
There is also the corresponding identity when one of the $F^{2m}$ has $K_2$ null,    
\begin{equation}
F^{2m}_{n:2:a+1}+F^{1m}_{n:a+2}
\longrightarrow 
F^{1m}_{n-1:1;a+2} \; . 
\label{eq:onemassidentity}
\end{equation}

The functional identities in eq.~(\ref{eq:twomassidentity}) and eq.~(\ref{eq:onemassidentity})  are very similar 
to those that appear in the collinear limit of the one-loop ``Maximally-Helicity-Violating" (MHV) amplitude in $\NeqFour$
although in that case the  identities are for the untruncated box integrals.   
The coefficients of the functions shown in \figref{fig:identity}, both descend to the appropriate coefficient in the collinear limit
\begin{equation}
c_{r+1,a+1}  \; , \;\; c_{r,a+2} \longrightarrow c_{r,a+2}
\end{equation}
and consequently
\begin{equation}
c_{r+1,a+1} F^{2m}_{n:r+1,a+1}+c_{r,a+2}F^{2m}_{n:r,a+2}  \longrightarrow c_{r,a+2} F^{2m}_{n-1:r,a+2}
\; . 
\end{equation}
This identifies where all the $F^{2m}$ terms of the $n-1$ point amplitude arise from in the collinear limit.

There are a few limiting cases in the $n$-point we must consider. 
The functions shown in~\figref{fig:vanishing} have vanishing coefficients and do not contribute.
The two functions represented in \figref{fig:boxdeltas} satisfy
\begin{equation}
F^{1m}_{n:a}+F^{1m}_{n:a+1} \longrightarrow -\Delta'
\end{equation}
and have coefficients satisfying
\begin{equation}
-{i  c_{1,a} \over 3 \spa{1}.2 \cdots \spa{n}.1}   \; , -{ i c_{1,a+1}  \over 3 \spa{1}.2 \cdots \spa{n}.1}  \longrightarrow S^{++,\rm tree}_{-}A^{(1)}_{n-1}  
\end{equation}
so that 
\begin{equation}
-{i  \over 3 \spa{1}.2 \cdots \spa{n}.1}\left(   c_{1,a} F^{1m}_{n:a} +c_{1,a+1}F^{1m}_{n:a+1}  \right)
\longrightarrow  -S^{++,\rm tree}_{-}A^{(1)}_{n-1} \Delta' 
\end{equation}
as required by eq.~(\ref{eq:collinearrequirement}).

\begin{figure}[H]
    \begin{picture}(450,100)(-50,0)
\CCirc(125,50){2}{Black}{Black}
\CCirc(120,57){2}{Black}{Black}
\CCirc(120,43){2}{Black}{Black}
\CCirc(265,50){2}{Black}{Black}
\CCirc(260,57){2}{Black}{Black}
\CCirc(260,43){2}{Black}{Black}
\Line(50,50)(75,75)
\Line(50,50)(75,25)
\Line(75,25)(100,50)
\Line(75,75)(100,50)
\Line(75,75)(75,100)
\Line(75,25)(75,0)
\Line(50,50)(25,50)
\Line(100,50)(125,70)
\Line(100,50)(125,30)
\Text(145,50)[c]{$+$}
\Line(190,50)(215,75)
\Line(190,50)(215,25)
\Line(215,25)(240,50)
\Line(215,75)(240,50)
\Line(215,75)(215,100)
\Line(215,25)(215,0)
\Line(190,50)(165,50)
\Line(240,50)(265,70)
\Line(240,50)(265,30)

\Text(95,100)[c]{$k_{a+1}$}
\Text(25,60)[c]{$k_{a}$}
\Text(170,60)[c]{$k_{a+1}$}
\Text(200,5)[c]{$k_{a}$}
\SetColor{Green}
\SetWidth{2}
\Line(50,50)(25,50)
\Line(75,75)(75,100)
\Line(190,50)(165,50)
\Line(215,25)(215,0)
\Text(315,50)[c]{\Large $\longrightarrow  -\Delta'$ }
\end{picture}    
\caption{The collinear limit of these two $F$-functions.}
\label{fig:boxdeltas}
\end{figure}
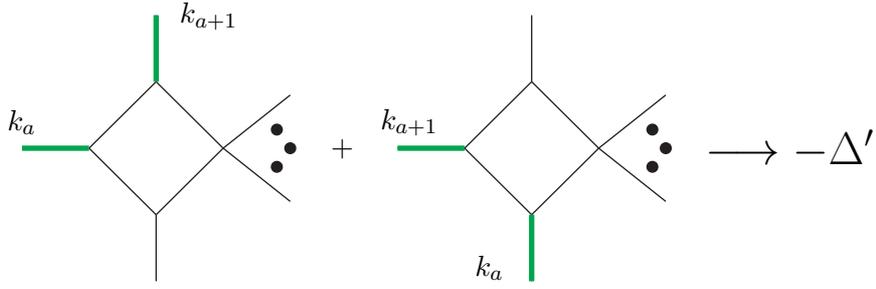

Consequently, the ansatz has the correct collinear limit up to rational terms.

\section{Unitarity Checks}

This section is rather a reverse of how the ansatz was formed.  We can use unitarity, subject to certain assumptions, to generate  the functions in the 
remainder function.  This is possible using four-dimensional unitarity where the cuts are evaluated in four dimensions rather that $4-2\eps$ dimensions. 
This process in principle will miss certain functions, specifically purely rational terms and terms that are sub-leading in $\epsilon$. 
As discussed above, we assume that the sub-leading in $\epsilon$ terms can be deduced from IR consistency allowing us to  generate an ansatz for the
polylogarithmic part of $F^{(2)}_n$.    

Using four dimensional unitarity, the one-loop all-plus amplitude has no cuts and may be regarded as a vertex.  
With this philosophy, the coefficient of the one-loop functions can be determined by one's favourite technique.  We use quadruple cuts~\cite{BrittoUnitarity} to determine  the box coefficients and canonical forms~\cite{Dunbar:2009ax} for the triangle and bubble functions. 
\begin{figure}[H]
\centerline{
    \begin{picture}(170,150)(-0,-50)    
     \Line( 0, 0)( 0,60)
     \Line( 0,60)(60,60)
     \Line(60,60)(60, 0)
     \Line(60, 0)( 0, 0)
     \Line( 0, 0)(0,-25)
     \Line( 0, 0)(-25,0)
     \Line( 0,60)(-15,75)
     \Line(60,60)(60,85)
     \Line(60,60)(85,60)
     \Line(60, 0)(68,-8)
   \CCirc(60,60){8}{Black}{Yellow} 
    \CCirc(0,0){8}{Black}{Yellow} 
    \Text(0,0)[c]{$A$}
     \Text(60,60)[c]{$B$}
     \CCirc(75,75){2}{Black}{Black} 
      \CCirc(81,68){2}{Black}{Black} 
      \CCirc(68,81){2}{Black}{Black} 
    \CCirc(-15,-15){2}{Black}{Black}
    \CCirc(-21,-8){2}{Black}{Black}
     \CCirc(-8,-21){2}{Black}{Black}
     \Text(22,-8)[c]{${}^-$}
      \Text(37,-8)[c]{${}^+$}
     \Text(22,62)[c]{${}^-$}
      \Text(37,62)[c]{${}^+$}  
     \Text(-30,0)[c]{${}^+$}   
     \Text(68,35)[c]{${}^+$}   
     \Text(68,22)[c]{${}^-$}  
     \Text(-8,35)[c]{${}^+$}   
     \Text(-8,22)[c]{${}^-$}   
     \Text(75,-18)[c]{${i-1}^+$}   
         \Text(10,-28)[c]{${i}^+$}   
                  \Text(-10,85)[c]{${i+r}^+$}   
      \Text(90,60)[c]{${}^+$}   
      \Text(60,90)[c]{${}^+$}   
\SetWidth{2}
     \DashLine(30,-10)(30,10){2}
          \DashLine(30,50)(30,70){2}
               \DashLine(-10,30)(10,30){2}     
               \DashLine(50,30)(70,30){2}
    \end{picture} 
    }
    \caption{The non-vanishing quadruple cut. $A$ is a MHV tree amplitude whereas $B$ is a one-loop all-plus amplitude.}
    \label{fig:oneloopstyle}
\end{figure}

When  all external legs are of positive helicity the non-vanishing quadruple cut shown in \figref{fig:oneloopstyle} is
\begin{align}
M_3&( (i-1)^+,-l_2^+,l_1^-)
\times
M_{r+2}^{\rm tree}( i^+, \cdots  ,(i+r-1)^+,-l_3^-,l_2^-)
\times
\notag \\
&M_3( (i+r)^+,-l_4^-,l_3^+) 
\times
M_{n-r}^{(1)}( (i+r+1)^+, \cdots ,(i-1)^+,-l_1^+,l_4^+) 
\end{align}
where the tree amplitude may be a three-point amplitude ($r=1$) but the one-loop amplitude must have at least two external legs. 
Using the $n$-point all-plus one-loop amplitude~\cite{Bern:1993qk}, 
\begin{align}
A^{(1)}(1^+,2^+,\cdots,n^+)=-{i\over 3}\sum_{1\leq k_1<k_2<k_3<k_4\leq n} 
{\spa{k_1}.{k_2} \spb{k_2}.{k_3}\spa{k_3}.{k_4}\spb{k_4}.{k_1} \over \spa{1}.2\spa{2}.3 \cdots\spa{n}.1}  
+O(\epsilon) \; ,  
\end{align}
we obtain
\begin{align}
&{ [i-1|K_4|i+r\ra 
{ [i+r|K_4|i-1\ra }    \over 
\spa1.2\spa2.3\spa3.4\cdots \spa{n}.1   }
\notag \\
& \hskip 2truecm\times
\left( 
\sum_{a<b<c<d \in K_4} \tr_{-}[abcd]-\sum_{a<b<c\in K_4} \tr_-[abc K_4]  +\sum_{a<b\in K_4} { \la i-1 |K_4  a b   K_4 | i+r\ra  \over \spa{i-1}.{i+r} }
\right) \; . 
\end{align}
This is the coefficient of the one-loop box function  $I_4^{2m}$.
This integral function satisfies
\begin{equation}
-2 (ST-K_2^2K_4^2) I_4^{2m}
= \left(  -{ (-S/\mu^2)^{-\epsilon} \over \eps^2 }
-{ (-T/\mu^2)^{-\epsilon} \over \eps^2 }
+{ (-K_2^2/\mu^2)^{-\epsilon} \over \eps^2 }
+{ (-K_4^2/\mu^2)^{-\epsilon} \over \eps^2 } \right)
+F^{2m}[S,T,K_2^2,K_4^2] 
\end{equation}  
where $[i-1|K_4|i+r\ra { [i+r|K_4|i-1\ra }=ST-K_2^2K_4^2$. This contains  
IR-infinite terms, $(-s/\mu^2)^{-\epsilon}/\eps^2$, together with $F^{2m}$.

\begin{figure}[H]
\centerline{    \begin{picture}(130,150)(0,-40)    
    \Line( 0, 0)( 60,0)
    \Line( 0, 0)(30,60)
     \Line(60,0)(30,60)
     \Line(0, 0)(-25, 0)     
     \Line(0, 0)(0,-25)
      \Line(60, 0)(85, 0)     
     \Line(60, 0)(60,-25)
      \Line(30, 60)(30,80)     
    \Text(40,75)[c]{$d^+$}  
    \Text(90,-20)[c]{$K_3$}  
    \Text(-30,-20)[c]{$K_1$}  
     \Text(38,-8)[c]{${}^+$}  
     \Text(22,-8)[c]{${}^-$}  
     \Text(7,27)[c]{${}^-$}  
     \Text(16,37)[c]{${}^+$}  
     \Text(53,27)[c]{${}^+$}  
     \Text(45,38)[c]{${}^-$}  
   \CCirc(60,0){8}{Black}{Yellow} 
   \CCirc(0,0){8}{Black}{Yellow} 
   \Text(60,0)[c]{$B$}
   \Text(0,0)[c]{$A$}
\Text(0,-35)[c]{$a^+$} 
\Text(60,-35)[c]{$f^+$} 
   \CCirc(-15,-15){2}{Black}{Black} 
   \CCirc(-8,-20){2}{Black}{Black} 
   \CCirc(-20,-8){2}{Black}{Black} 
   \CCirc(75,-15){2}{Black}{Black} 
   \CCirc(68,-20){2}{Black}{Black} 
   \CCirc(80,-8){2}{Black}{Black} 
 \SetWidth{2}
 \DashLine(30,-8)(30,8){2}
 \DashLine(20,26)(10,34){2}
  \DashLine(40,26)(50,34){2}
    \end{picture} 
    \begin{picture}(130,150)( 0,-20)  
     \Line(10,10)(10,70) 
     \Line( 0,75)(65,40)
     \Line( 85,60)(65,40)
      \Line( 85,20)(65,40)
      \Line(0,5)( 65,40)
    \CCirc(65,40){8}{Black}{Yellow}       
    \Text(65,40)[c]{$B$}
    \CCirc(90,40){2}{Black}{Black}    
    \CCirc(85,50){2}{Black}{Black}    
    \CCirc(85,30){2}{Black}{Black}    
\Text(-8,5)[c]{${e}^+$}   
\Text(-8,75)[c]{${f}^+$}   
\Text(85,65)[c]{${}^+$}   
\Text(85,10)[c]{${}^+$}   
\Text(45,55)[c]{${}^+$}   
\Text(45,20)[c]{${}^+$}   
\Text(35,62)[c]{${}^-$}   
\Text(35,13)[c]{${}^-$}   
\Text(105,40)[c]{$K_3$}
    \Text(3,45)[c]{${}^{+}$}      
    \Text(3,33)[c]{${}^{-}$}  
        \Text(17,45)[c]{${}^{-}$}      
    \Text(17,33)[c]{${}^{+}$}  
     \SetWidth{2}
 \DashLine(0,40)(20,40){2}
 \DashLine(35,51)(44,66){2}
 \DashLine(35,24)(44,15){2}
  \end{picture} 
    }
    \caption{The non-vanishing triple cuts. The one-mass triangle has two non-vanishing configuration. $A$ and $B$ are as in \figref{fig:oneloopstyle}.}
    \label{fig:triplecuts}
\end{figure}
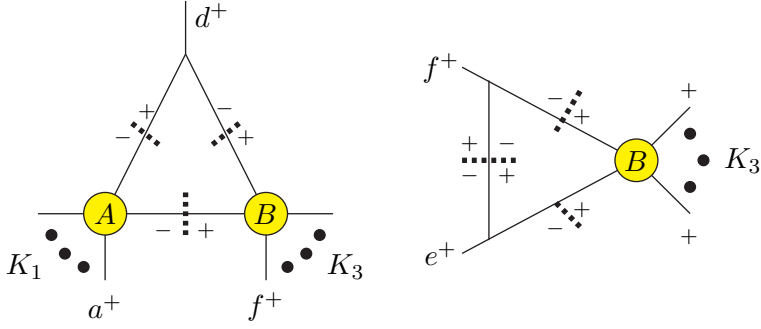
There are also triangle integral functions whose coefficient may be determined from 
triple cuts~\cite{Bidder:2005ri,Darren,BjerrumBohr:2007vu,Mastrolia:2006ki}.
The non-vanishing triple cuts as shown in~\figref{fig:triplecuts}. 
These can be evaluated using canonical forms~\cite{Dunbar:2009ax}. The coefficient of the two mass triangle shown in~\figref{fig:triplecuts} is
\begin{equation}
{\spa{f}.a [d | K_3 |d \ra  \over  \spa{1}.2\spa{2}.3 \cdots\spa{n}.1   \spa{a}.d  \spa{f}.d  }
\sum_{ i < j \in K_3} \la d | K_3 i j K_3|d \ra 
\end{equation}
and that of  the one-mass triangle is
\begin{align}
{   s_{ef} \spa{d}.e \spa{e}.f \sum_{i<j\in K_3} \spb{e}.i   \spa{i}.j \spb{j}.{e}   \over \spa{1}.2\spa{2}.3 \cdots\spa{n}.1  \spa{d}.f  } 
+
{ s_{ef}  \spa{e}.f \spa{a}.f   \sum_{i<j\in K_3}   \spb{f}.i   \spa{i}.j \spb{j}.{f}\over \spa{1}.2\spa{2}.3 \cdots\spa{n}.1 \spa{e}.{a}   } 
\; . 
\end{align}
The integral functions are
\begin{align}
I_{3}^{2 \rm m}\bigl(K_1^2,K_3^2\bigr)= {1\over\e^2}
{(-K_1^2/\mu^2)^{-\e}-(-K_3^2/\mu^2)^{-\e} \over  (-K_1^2)-(-K_3^2) }
\end{align}
and
\begin{align}
I_{3}^{1\rm m}(K^2_3) = {1\over\e^2} (-K^2_3/\mu^2)^{-1-\e}\ .
\end{align}
Each IR divergent term, such as $ (-S/\mu^2)^{-\e}/\e^2$, occurs in both triangle and the box functions. When we sum over the box and triangle 
contributions we obtain an
overall coefficient of $A^{(1),\epsilon^0}_n(1^+,2^+,\cdots ,n^+)$,
\begin{align}&
\left(   \sum  {\cal C}_{i} I_{4,i}^{\rm 2m}
+\sum  {\cal C}_{i}  I_{3,i}^{2 \rm m}  
+\sum  {\cal C}_{i}  I_{3,i}^{1 \rm m}  \right) \biggl|_{IR}
= 
A^{(1),\epsilon^0}_n(1^+,2^+,\cdots, n^+)
\times \sum_{i=1}^{n} \frac{1}{\epsilon^2} \left(\frac{\mu^2}{-s_{i,i+1}}\right)^{\epsilon} ,
\end{align}
where $A^{(1),\epsilon^0}_5(1^+,2^+,\cdots, n^+)$ is the order $\epsilon^0$ truncation of the one-loop amplitude\footnote{This has been checked on random kinematic points for $n\leq 25$.}.
A key step is to promote the coefficient of these terms to be the all-$\epsilon$ form of the one-loop amplitude which 
then gives the correct singular structure of the amplitude. 

Note that in principle the amplitude could contain bubble functions. These can be obtained from two-particle cuts. However, the two-particle cuts
are ${\cal O}(\ell^{-1})$, indicating that the bubble functions have vanishing coefficients and are thus absent. This is consistent with the known
IR and UV singular structure of these particular amplitudes. 

\section{Rational Terms}

The rational terms in the amplitude are of course very important.    After identifying the non-rational part the rational part may be obtained
 by recursion. This was illustrated for the five point case in~\cite{Dunbar:2016aux}.   Recursion is however
fairly complicated because the rational terms 
contain double poles  which means we require subleading information about the amplitude.  

While this has proved possible for some amplitudes~\cite{Dunbar:2010xk,Alston:2012xd,Alston:2015gea,Dunbar:2016dgg},
we have not yet been able to obtain the rational part of the $n$-point amplitude in closed form.

\section{Conclusions} 

We have proposed an explicit compact expression for the polylogarithms  of the all-plus two-loop $n$-point amplitude which has:

a) the correct IR and UV structure

b) the correct collinear limits

c) the correct four dimensional cuts

\noindent 
A key element is that the most complex polylogarithms are contained in the leading singular terms. 
The expression for the remaining polylogarithms is constructed from simple building blocks which are combinations of dilogarithms corresponding to 
simple one-loop box integrals.  
Whether there is an underlying symmetry reason for the 
simplicity (such as the link between the one-loop all-plus amplitude and the 
amplitudes of self-dual Yang-Mills~\cite{Cangemi:1996rx,Chalmers:1996rq}) remains to be seen.
We hope that the 
simplicity of the result
will inspire attempts to produce compact analytical expressions for further multi-loop gauge theory amplitudes.

\section{Acknowledgements}

This work was supported by STFC grant ST/L000369/1.


\begin{thebibliography}{99}
\bibitem{Eden}
R.J. Eden, P.V. Landshoff, D.I. Olive, J.C. Polkinghorne, {\it
The Analytic S Matrix}, (Cambridge University Press, 1966).


\bibitem{Bern:1994zx}
  Z.~Bern, L.~J.~Dixon, D.~C.~Dunbar and D.~A.~Kosower,
  Nucl.\ Phys.\ B {\bf 425} (1994) 217
  [hep-ph/9403226].
  
  

\bibitem{Bern:1994cg}
  Z.~Bern, L.~J.~Dixon, D.~C.~Dunbar, D.~A.~Kosower,
  Nucl.\ Phys.\  {\bf B435 } (1995)  59
  [hep-ph/9409265].
        




\bibitem{Britto:2005fq}
  R.~Britto, F.~Cachazo, B.~Feng and E.~Witten,
  Phys.\ Rev.\ Lett.\  {\bf 94} (2005) 181602
  [hep-th/0501052].

\bibitem{Badger:2013gxa}
  S.~Badger, H.~Frellesvig and Y.~Zhang,
  JHEP {\bf 1312} (2013) 045
  doi:10.1007/JHEP12(2013)045
  [arXiv:1310.1051 [hep-ph]].

\bibitem{Badger:2015lda}
  S.~Badger, G.~Mogull, A.~Ochirov and D.~O'Connell,
  JHEP {\bf 1510} (2015) 064
  doi:10.1007/JHEP10(2015)064
  [arXiv:1507.08797 [hep-ph]].



\bibitem{Gehrmann:2015bfy}
  T.~Gehrmann, J.~M.~Henn and N.~A.~Lo Presti,
  Phys.\ Rev.\ Lett.\  {\bf 116} (2016) 6,  062001
  doi:10.1103/PhysRevLett.116.062001
  [arXiv:1511.05409 [hep-ph]].

\bibitem{Dunbar:2016aux}
  D.~C.~Dunbar and W.~B.~Perkins,
  arXiv:1603.07514 [hep-th].



\bibitem{Catani:1998bh}
  S.~Catani,
  Phys.\ Lett.\ B {\bf 427} (1998) 161
  doi:10.1016/S0370-2693(98)00332-3
  [hep-ph/9802439].
 
 

\bibitem{Bern:1993qk}
  Z.~Bern, G.~Chalmers, L.~J.~Dixon and D.~A.~Kosower,
  Phys.\ Rev.\ Lett.\  {\bf 72} (1994) 2134
  doi:10.1103/PhysRevLett.72.2134
  [hep-ph/9312333].
 


\bibitem{Bern:1996ja}
  Z.~Bern, L.~J.~Dixon, D.~C.~Dunbar and D.~A.~Kosower,
  Phys.\ Lett.\ B {\bf 394} (1997) 105
  doi:10.1016/S0370-2693(96)01676-0
  [hep-th/9611127].
  


  
\bibitem{BrittoUnitarity} R.~Britto, F.~Cachazo and B.~Feng,
  Nucl.\ Phys.\ B {\bf 725} (2005) 275 [hep-th/0412103].

\bibitem{Bidder:2005ri}
  S.~J.~Bidder, N.~E.~J.~Bjerrum-Bohr, D.~C.~Dunbar and W.~B.~Perkins,
  Phys.\ Lett.\  B {\bf 612} (2005) 75
  [hep-th/0502028].
  
  

\bibitem{Bern:2002tk}
  Z.~Bern, A.~De Freitas and L.~J.~Dixon,
  JHEP {\bf 0203} (2002) 018
  doi:10.1088/1126-6708/2002/03/018
  [hep-ph/0201161].


\bibitem{Dunbar:2009ax}
  D.~C.~Dunbar, W.~B.~Perkins and E.~Warrick,
  JHEP {\bf 0906} (2009) 056
  [arXiv:0903.1751 [hep-ph]].
  



\bibitem{Darren}
  D.~Forde,
  Phys.\ Rev.\ D {\bf 75} (2007) 125019
  [arXiv:0704.1835 [hep-ph]].

\bibitem{BjerrumBohr:2007vu}
  N.~E.~J.~Bjerrum-Bohr, D.~C.~Dunbar and W.~B.~Perkins,
  JHEP {\bf 0804} (2008) 038
  [arXiv:0704.1835 [hep-ph]].



\bibitem{Mastrolia:2006ki}
  P.~Mastrolia,
  Phys.\ Lett.\  B {\bf 644} (2007) 272
  [arXiv:hep-th/0611091].
  


  


  

\bibitem{Dunbar:2010xk}
  D.~C.~Dunbar, J.~H.~Ettle and W.~B.~Perkins,
  JHEP {\bf 1006} (2010) 027
  [arXiv:1003.3398 [hep-th]].


\bibitem{Alston:2012xd}
  S.~D.~Alston, D.~C.~Dunbar and W.~B.~Perkins,
  Phys.\ Rev.\ D {\bf 86} (2012) 085022
  [arXiv:1208.0190 [hep-th]].
  
\bibitem{Alston:2015gea}
  S.~D.~Alston, D.~C.~Dunbar and W.~B.~Perkins,
  Phys.\ Rev.\ D {\bf 92} (2015) 6,  065024
  doi:10.1103/PhysRevD.92.065024
  [arXiv:1507.08882 [hep-th]].

\bibitem{Dunbar:2016dgg}
  D.~C.~Dunbar and W.~B.~Perkins,
  arXiv:1601.03918 [hep-th].


\bibitem{Cangemi:1996rx}
  D.~Cangemi,
  Nucl.\ Phys.\ B {\bf 484} (1997) 521
  doi:10.1016/S0550-3213(96)00586-X
  [hep-th/9605208].

\bibitem{Chalmers:1996rq}
  G.~Chalmers and W.~Siegel,
  Phys.\ Rev.\ D {\bf 54} (1996) 7628
  doi:10.1103/PhysRevD.54.7628
  [hep-th/9606061].


\end{thebibliography}
\end{document}